\documentclass[12pt]{iopart}

\usepackage[numbers, sort&compress]{natbib} 
\usepackage{hyperref}

\hypersetup{
citecolor=blue,
linkcolor=blue,
colorlinks=true,
}
\usepackage{graphicx}

\begin{document}

\title[]{Surfactant-Assisted Optothermal Binding of Plasmonic Colloids at a Droplet Interface}

\author{Sumant Pandey, G V Pavan Kumar}

\address{Department of Physics, Indian Institute of Science Education and Research (IISER) Pune, Pune, 411008, India}

\ead{pavan@iiserpune.ac.in}
\vspace{10pt}
\begin{indented}
\item[]
\end{indented}

\begin{abstract} 
Light-driven assembly of plasmonic colloids at liquid–liquid interfaces provides a route to dynamically organize colloidal matter beyond conventional optical trapping. While previous studies of optical binding have primarily explored scattering-mediated particle interactions, absorption-dominated optothermal assembly of plasmonic nanoparticles at liquid–liquid interfaces remains comparatively unexplored. Here, we demonstrate low-power, 532 nm laser-induced optothermal trapping and reversible assembly of 400 nm gold nanoparticles at a surfactant-stabilized bromododecane droplet/water interface. Surfactant-assisted optothermal recruitment leads to the sequential formation of optothermally bound assemblies, evolving from a single trapped particle to dimers, trimers, quadramers, and occasionally pentamers. Their spatiotemporal dynamics reveal micrometre-scale nearest-neighbour separations, continuous structural reorganization, and dynamic mobility. Switching off the laser leads to disassembly and recovery of Brownian diffusion, demonstrating reversible optical control. Control experiments establish that sustained particle recruitment requires the combined presence of the liquid–liquid interface, surfactant, and optical excitation, consistent with coupled photothermal, thermophoretic, and hydrodynamic transport. Notably, comparable nearest-neighbour separations are observed for 200 and 400 nm gold nanoparticles, indicating that the characteristic spacing is largely insensitive to particle size over the investigated range. These findings establish surfactant-stabilized liquid–liquid interfaces as a platform for absorption-driven, reversible organization of plasmonic matter and for investigating nonequilibrium collective dynamics, with potential relevance to nanoparticle assembly at biological interfaces such as cell surfaces.
\end{abstract}

%

%
%
%

\section{Introduction}
Optical forces arise from the exchange of momentum between light and matter, enabling contactless trapping and manipulation of microscopic objects \cite{jones2015optical}. Ashkin and co-workers pioneered light-driven particle manipulation through radiation-pressure trapping and subsequently demonstrated optical confinement of atoms and dielectric particles, establishing the foundation of modern optical tweezers \cite{ashkin1970atomic,chu1985three,chu1986experimental}. Optical trapping was subsequently extended to biological systems, including viruses and bacteria \cite{ashkin1987optical}. Today, optical tweezers provide versatile control over colloids, nanoparticles, molecular systems, and soft matter, enabling controlled trapping, transport, and collective assembly \cite{grier2003revolution,dienerowitz2008optical,volpe2023roadmap,dholakia2010colloquium,forbes2020optical}. Optical confinement has also been employed to probe the fluctuations and active forces experienced by passive tracer particles in living cells \cite{nakul2023studying}.

The optical manipulation of plasmonic colloids is attractive because metallic nanoparticles exhibit strong, wavelength-dependent absorption and scattering, enabling light-mediated forces and interparticle interactions \cite{marago2013optical,tong2011plasmon,odebo2020optical}. Plasmonic gold nanoparticles also provide strong electromagnetic-field enhancement for applications such as ultrasensitive Raman detection \cite{chaubey2024ultrasensitive}. Optical binding provides a route to nanoparticle organization through light-induced interparticle forces, with configurations governed by particle properties, separation, polarization, and beam characteristics \cite{bradshaw2005optically,thirunamachandran1980intermolecular,rodriguez2008optical,vcivzmar2010multiple}. Recent studies have further demonstrated rotational dynamics and collective organization of optically bound nanoparticle assemblies under single and multiple laser beams \cite{tao2023rotational,tao2024generalized} highlighting the collective and nonequilibrium dynamics of optically driven colloidal systems \cite{martinez2016colloidal, cerbino2017perspective}. Experiments have also demonstrated optical binding and light-controlled assembly of metallic nanoparticles \cite{demergis2012ultrastrong,yan2013guiding,yan2014potential,kudo2018single}. However, optical absorption also produces photothermal heating, which can modify the surrounding medium and particle dynamics. Optothermal effects can therefore become an important component of plasmonic colloid manipulation, with thermally induced forces contributing alongside direct optical interactions \cite{setoura2019opto,rubinsztein1998optical,baffou2013thermo,kumar2022plasmonic}.

Light-induced heating provides an alternative route for particle manipulation by generating thermal gradients and associated thermophoretic, thermo-osmotic, thermoelectric, and hydrodynamic flows \cite{chen2022heat,piazza2008thermophoresis,kollipara2023optical,baffou2020applications,bregulla2016thermo,franzl2022hydrodynamic,wurger2010thermal}. Thermophoresis provides a general mechanism for particle transport in non-uniform temperature fields, while thermoplasmonic heating enables localized manipulation using metallic nanoparticles as nanoscale heat sources \cite{piazza2008thermophoresis,baffou2020applications,kollipara2023optical}. 
These optothermal effects have enabled particle trapping, transport, assembly, and photothermally driven flows in plasmonic and colloidal systems, with thermo-fluid dynamics playing an important role in optical manipulation under confined conditions \cite{roxworthy2014understanding,paul2022optothermal,tiwari2021single,chand2025optothermal,jin2018photothermal,tsuji2022optical}. Plasmon-enhanced thermophoresis has further enabled low-power trapping and reversible assembly of plasmonic nanoparticles, as well as collective organization of colloidal matter \cite{lin2016light,lin2017opto,lin2018opto,braun2013optically}. In surfactant-containing systems, thermal gradients can couple to surfactant and micelle redistribution, generating thermoelectric and concentration gradients that modify colloidal transport and trapping \cite{vigolo2010thermophoresis,lin2018opto,tiwari2021single,kollipara2023optical,chen2021liquid}. Photothermally induced Marangoni flows have also been exploited to direct colloidal particles into patterns, control nanoparticle deposit morphology, and achieve optothermofluidic particle trapping \cite{varanakkottu2016light,anyfantakis2017evaporative,ragisha2024moving}. Related studies have further demonstrated optothermally driven particle transport and rotational motion at fluid interfaces \cite{zhang2025optothermal,meng2025orbiting,wang2011micro}. Optothermal manipulation has also been demonstrated near droplet surfaces, where laser-induced Marangoni convection and capillary forces drive the oscillatory motion of hollow microparticles \cite{zhang2025optothermal}. More recently, ice-phase optothermal tweezers have demonstrated programmable manipulation and assembly of metallic and dielectric nanoparticles at a laser-generated ice–water interface, extending optothermal control to frozen environments \cite{zhou2026ice}. Together, these studies highlight the versatility of optothermal manipulation across fluid and phase-change interfaces.

At liquid-liquid interfaces, optothermal effects couple to interfacial tension, fluid motion, and interfacial mechanics, providing a soft and deformable environment for nanoparticle transport and assembly \cite{bickel2013flow,bregulla2016thermo,franzl2022hydrodynamic,wang2011micro}. They can also serve as model systems to study biological surfaces such as cells. Unlike rigid solid-liquid interfaces, fluid interfaces offer weak, tunable confinement that preserves lateral nanoparticle mobility, enabling adsorption, diffusion, transport, and reorganization \cite{lin2003nanoparticle,shi2018nanoparticle}. Interfacial adsorption can promote nanoparticle localization, while the fluid boundary permits dynamic rearrangement and diverse interfacial assemblies \cite{lin2003nanoparticle,shi2018nanoparticle}. Nanoparticle self-assembly has also been demonstrated at liquid-crystal droplet interfaces, highlighting the role of interfacial organization and interactions in directing nanoparticle assembly at curved fluid boundaries \cite{rahimi2015nanoparticle}. 

Conventional colloidal assembly at liquid-liquid interfaces has largely focused on spontaneous adsorption and interparticle interactions, where particles become strongly localized at the interface and can form ordered two-dimensional structures \cite{mcgorty2010colloidal,bresme2007nanoparticles,hu2012oil}. Such assemblies are often stabilized by substantial interfacial adsorption energies, capillary interactions, ligand-mediated forces, or specific molecular linkers and are therefore typically difficult to reversibly reorganize \cite{lin2003nanoparticle,shi2018nanoparticle}. Chemical interactions have also been used to control the order and geometry of plasmonic patches at soft curved interfaces \cite{chen2026chemically}. Similarly, interfacial confinement has enabled the fabrication of metal-coated Janus colloids through electroless deposition \cite{yele2026janus}. DNA-mediated strategies further enable programmable assembly of plasmonic and colloidal nanoparticles through specific interparticle binding \cite{fan2011dna,rogers2016using}. Optofluidic crystallisation and interfacial optical trapping have demonstrated long-range particle transport and collective colloidal organization at fluid interfaces, primarily involving tethered dielectric colloidal particles, rather than freely diffusing plasmonic nanoparticles at a droplet interface. \cite{caciagli2020controlled,tsuboi2021optical}. Previous studies have also reported dielectric nanoparticle self-assembly at liquid-crystal droplet interfaces \cite{rahimi2015nanoparticle}, optothermal oscillation of hollow microparticles near droplet surfaces \cite{zhang2025optothermal}, and optothermal assembly at ice–water interfaces \cite{zhou2026ice}. Our recent work demonstrated the optical-tweezer-assisted coupling of liquid-crystal microdroplets with drop-cast plasmonic gold nanoparticle clusters, enabling reversible control of whispering-gallery modes \cite{pandey2026optical}. 
While previous studies of optical binding have explored scattering-mediated particle interactions, the assembly dynamics of plasmonic nanoparticles under absorption-dominated excitation remain comparatively unexplored at liquid–liquid interfaces. Here, we have shown under low-power, continuous 532 nm laser-induced optothermally bounded and sequential assembly of freely diffusing gold nanoparticles at a surfactant-stabilized oil–water droplet interface. 

In this work, we demonstrate visible-light-induced optothermal trapping and reversible assembly of plasmonic gold nanoparticles at a surfactant-stabilized oil–water droplet interface under continuous 532 nm laser illumination, as schematically illustrated in Fig. 1(a). Unlike conventional interfacial assembly approaches that rely primarily on spontaneous adsorption or irreversible particle attachment, our approach enables the dynamic recruitment and reorganization of freely diffusing plasmonic nanoparticles near the illuminated droplet interface. The assemblies exhibit several distinctive features: (i) stable interparticle separations (nearest neighbor of approximately 1.1 to 1.3 \(\mu\)m) of the order of micrometers; (ii) sequential evolution from dimers to higher-order clusters; (iii) reversible assembly and disassembly demonstrated through laser on/off experiments; (iv) control experiments establishing that sustained particle recruitment requires the combined presence of the liquid–liquid interface, SDS, and optical excitation; and (v) comparable characteristic interparticle separations for 200 and 400 nm gold nanoparticles over the investigated size range. These observations reveal that optothermal assembly at liquid–liquid interfaces is governed by the interplay of photothermal transport, interfacial hydrodynamics, optical interactions, and particle–particle interactions. Our findings establish surfactant-stabilized liquid–liquid interfaces as a dynamic platform for reversible, light-driven assembly and nonequilibrium dynamics of plasmonic matter, with potential relevance to nanoparticle organization at biological interfaces.

\begin{figure}[h!]
\centering
\includegraphics[width=\linewidth]{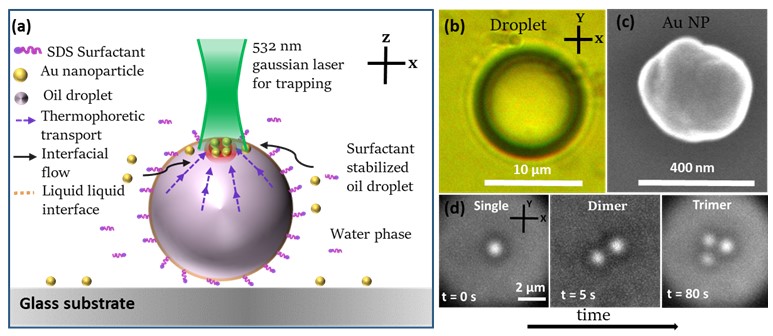}
\caption{Experimental configuration and optothermal assembly of plasmonic colloids at a liquid-liquid interface.
(a) Schematic of the experimental configuration showing a bromododecane (BDD) droplet at the BDD-water interface stabilized by sodium dodecyl sulfate (SDS) and locally illuminated with a continuous-wave 532 nm laser. Localized optical heating at the interface drives optothermal recruitment and confinement of plasmonic colloids near the illuminated region. (b) Representative top-view bright-field image of a BDD droplet at the liquid–liquid interface, showing its approximately circular geometry (scale bar: 10~$\mu$m). (c) Scanning electron microscopy (SEM) image of a representative gold nanoparticle (AuNP) used in the experiments, with a diameter of approximately 400 nm (scale bar: 400 nm). (d) Representative bright-field time-lapse images showing the sequential evolution of the optothermal assembly from a single trapped plasmonic colloid to dimer and trimer configurations at the droplet interface under continuous 532 nm illumination. The complete assembly sequence is provided in Supplementary Video 1.}
\label{fig:figure01}
\end{figure}

\section{Experimental Section}
\subsection{Materials}
We have purchased gold nanoparticles of different diameters and sodium dodecyl sulfate (SDS) as an anionic surfactant and bromododecane (BDD) from Sigma-Aldrich. Deionised water (resistance \( \geq 18.2\  M\Omega\cdot cm\)) obtained from the Milli-Q water purification system was used in all of the experiments. The gold nanoparticles have diameters of 200 nm and 400 nm purchased from Sigma-Aldrich. 

\subsection{Sample Preparation}
We have prepared emulsions of isotropic oil bromododecane (BDD) in MQ water. For this experiment, approximately 3 \(\mu\)L of bromododecane (BDD) and 0.085\% w/w (3 mM) sodium dodecyl sulfate was used and mixed using a vortexer.
The mixing was performed using a Tarsons Spinix Vortex Shaker (Model No. 3020) operating at a speed of 3000 rpm for 60 seconds. These parameters were used to obtain droplets of a suitable size range for our experiments. Because BDD is an isotropic organic oil immiscible with water, the droplets remain stable throughout the experiment, with negligible changes in droplet size\cite {jonavs2017thermal}. Previous interfacial optofluidic crystallization studies have employed approximately 2.5 mM SDS (critical-micelle concentration) for colloidal assembly at fluid interfaces \cite{caciagli2020controlled}. Based on this interfacial precedent, we used approximately 3 mM SDS to stabilize the BDD–water interface and facilitate optothermal assembly of Au nanoparticles at the droplet surface. Suspensions of 400 nm diameter gold nanoparticles were prepared in water. After droplet formation in a 1.5 mL Eppendorf tube, a desired volume of the aqueous Au nanoparticle suspension was added to the tube, followed by shaking for a few seconds. 
Goldseal Cover Glass from Ted Pella, Inc., was used for experiments after cleaning with acetone. A double-sided adhesive spacer from Grace Bio-Laboratories (SecureSeal imaging spacer, 120 \(\mu\)m thickness) was used to build a microfluidic chamber. The spacer is pasted on the cover glass. 
About 15 \(\mu\)L of mixture solution is enclosed in a sample chamber between two glass coverslips. This method produces distinct, well-separated microdroplets scattered throughout the fluid medium. The micro-droplets of a suitable size range are highly stationary after a few hours of sample preparation. The chamber is placed on the stage of our optical trapping microscope. All observations were carried out at room temperature.

\subsection{Optical Measurements}
A 532 nm unpolarized Gaussian laser beam was used for optical trapping and optothermal assembly. The particles were trapped in upright geometry and imaged in brightfield mode. A 100×, 0.95 NA objective lens (Olympus MPlanApo N) focused the laser from the top of the AuNP anchored on the cover glass and captured images at 100 frames per second. The laser beam was focused near the liquid–liquid interface through a high numerical aperture microscope objective. Scattered light and nanoparticle trajectories were recorded using a CCD camera coupled to an optical microscope. The sample was illuminated with white light from the bottom. The laser wavelength was filtered from the collected light before reaching the camera using a notch filter. The setup was equipped with a high-speed, low-profile motorised scanning stage (MLS203-1) having dimensions of 260 mm \(\times\)\ 230 mm \(\times\)\ 31 mm, translation stage for high-precision control of the sample. The laser powers used for trapping in the sample plane are about 6 mW.  The beam was not expanded, and the back aperture of the objective was underfilled. A schematic of the experimental setup is shown in Supplementary Fig. S1.

\subsection{Particle Tracking and Trajectory Analysis}

Time-resolved bright-field images were analysed using ImageJ for particle identification and tracking. The extracted particle coordinates were further analyzed using MATLAB to obtain individual trajectories and quantify the assembly dynamics. Interparticle separations, distance distributions and radial displacements were calculated from the time-resolved particle coordinates. These analyses were used to characterize the structural evolution and dynamical reorganization of the plasmonic-colloid assemblies.

\section{Results and discussion}

\subsection{ Optothermally Driven Assembly of Plasmonic Colloids at a Liquid–Liquid Interface }

Gold nanoparticles (AuNPs) were dispersed in an aqueous sodium dodecyl sulfate (SDS) solution, and bromododecane (BDD) droplets were introduced to form a liquid–liquid interface between the oil and aqueous phases. The SDS-stabilized BDD–water interface provides the interfacial environment in which plasmonic colloids can be optothermally confined and assembled. We investigated the formation and evolution of these assemblies under continuous-wave 532 nm laser illumination using an optical microscopy platform equipped with a high-numerical-aperture objective and a camera for time-resolved imaging. The experimental geometry and physical picture of the optothermal assembly process are illustrated schematically in Fig. 1a, while the detailed optical microscope configuration used for the experiments is provided in Supplementary Fig. S1.

As shown in Fig. 1a, the 532 nm Gaussian laser is focused at the apex of a BDD droplet resting on a glass substrate, locally illuminating the BDD–water interface. To characterize the optical response of the components involved in the experiment, we recorded UV–visible extinction spectra of the aqueous medium, BDD, and AuNP dispersions (Supplementary Fig. S2 and S3). The AuNP dispersion exhibits a broad extinction band centred at approximately 590 nm, providing substantial spectral overlap with the 532 nm excitation. For the 400 nm AuNPs, the measured extinction contains contributions from both absorption and scattering and therefore does not directly quantify the absorbed optical power or local temperature rise. Nevertheless, the spectral overlap confirms a strong optical response of the plasmonic colloids at the excitation wavelength, supporting localized photothermal heating under continuous illumination. At the liquid–liquid interface, such localized heating can establish thermal gradients that couple to interfacial transport processes and drive nonequilibrium colloidal motion \cite{caciagli2020controlled}. 

Under the experimental conditions, 3 mM SDS is present in the aqueous phase, above the reported critical micelle concentration. In the presence of both the SDS-stabilized liquid–liquid interface and optical illumination, freely diffusing plasmonic colloids are recruited toward the illuminated region, where they become confined and subsequently form multi-particle assemblies (Supplementary Videos 1). The observed recruitment is consistent with a coupled interfacial optothermal process involving localized heating, thermophoretic and hydrodynamic transport, and particle–interface interactions, with colloidal interactions contributing to the resulting spatial organization \cite{caciagli2020controlled}. Because the relative contributions of these mechanisms cannot be separated from the present measurements, we refer to the overall process as optothermal recruitment and confinement rather than assigning the particle motion to a single dominant mechanism. The optical image in Fig. 1b shows a representative BDD droplet with a characteristic diameter of approximately 10~$\mu$m; droplets with diameters of approximately 10–20~$\mu$m were used throughout the experiments. The assemblies occupy only a localized portion of the interface, allowing their lateral trajectories and structural configurations to be analysed in a quasi-two-dimensional geometry. The SEM image in Fig. 1c confirms that the plasmonic colloids are approximately spherical with a characteristic diameter of approximately 400 nm and relatively limited size dispersion. Figure 1d shows the initial stages of particle recruitment and assembly under continuous illumination, providing an overview of the evolution that is analysed quantitatively in the following sections (Supplementary Video 1). Together, Fig. 1 establishes the experimental conditions for optothermal manipulation of plasmonic colloids at the BDD–water interface. The combination of plasmonic optical response, localized 532 nm excitation, an SDS-stabilized liquid–liquid interface, and interfacial optothermal transport provides the basis for the sequential and reconfigurable assembly investigated below.

\subsection{Sequential Evolution of Plasmonic-Colloid Assemblies from a Single Particle to a Pentamer} 
The representative Figure 2 presents the time-lapse images illustrating the sequential evolution of the optothermal assembly under continuous 532 nm laser illumination. Starting from a single trapped plasmonic colloid, the assembly grows through successive particle recruitment, progressing from a single particle to dimer, trimer, quadramer, and, occasionally, pentamer configurations. Rather than undergoing rapid and irreversible aggregation, the assembly develops through discrete transitions separated by finite residence times. A second plasmonic colloid is recruited approximately 5 s after the initial trapping event, forming a dimer, followed by trimer formation after approximately 80 s and subsequent evolution to quadramer and pentamer configurations (Supplementary Video 1). These discrete intermediate states demonstrate that assembly growth proceeds through sequential particle recruitment, with each configuration being dynamically maintained for a finite period before transitioning to a larger assembly. The assembly size is spatially constrained by the localized optothermal interaction region generated at the liquid-liquid interface. Under the fixed laser intensity and focal conditions, plasmonic colloids within this effective interaction region can be recruited and confined, whereas particles approaching from outside the region are not stably retained and continue to undergo Brownian diffusion. The focused laser spot at the interface has a characteristic diameter of approximately 1.5~$\mu$m, providing localized optical excitation for the assembly. Quadramer formation is frequently observed and therefore provides a suitable configuration for quantitative trajectory analysis in the following section. Pentamer formation occurs less frequently, with the fifth colloid occasionally exhibiting reduced image intensity and a slight positional displacement, consistent with its location toward the periphery of the assembly or partial displacement from the focal plane. Nevertheless, the pentamer represents the largest transient configuration observed under the present experimental conditions. Throughout the growth process, the constituent plasmonic colloids remain dynamically mobile and undergo positional fluctuations within the assembly rather than becoming irreversibly fixed. Thus, sequential particle recruitment produces reconfigurable nonequilibrium configurations that are continuously sustained by the applied optothermal field.    

\begin{figure}[h!]
  \centering
  \includegraphics[width=\linewidth]{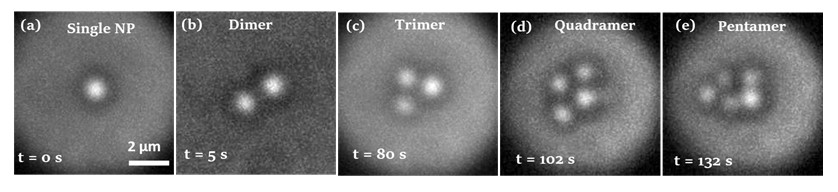}
  \caption{Sequential optothermal assembly of plasmonic colloids at a liquid-liquid interface.
  Time-resolved bright-field microscopy images showing the sequential evolution of plasmonic colloids under continuous 532 nm laser illumination. Starting from a single trapped colloid, successive particle recruitment leads to the formation of a dimer (5 s), trimer (80 s), quadramer (102 s), and pentamer (132 s). The assembled colloids remain dynamically confined near the illuminated interface while undergoing positional fluctuations and structural reorganization. Supplementary Video 1 shows the complete assembly sequence. Scale bar: 2~$\mu$m.}
  \label{fig:figure02}
\end{figure}

\subsection{ Reversible Optothermal Trapping and Disassembly }
To demonstrate the reversible nature of the optothermal assembly, we monitored the nanoparticle configuration during repeated laser on/off cycles (Fig. 3; Supplementary Video 2, presented at 2$\times$ playback speeds). Under continuous 532 nm illumination, three Au nanoparticles form and remain confined in a trimer-like configuration at the liquid-liquid interface. When the laser is switched off at $t=4$ s, the assembly rapidly loses confinement, and the nanoparticles begin to diffuse away from the illuminated region. Upon restoring the laser illumination at $t=10$ s, a single nanoparticle is first trapped at the illuminated region, followed by the sequential recruitment of a second nanoparticle at $t=53$ s and a third nanoparticle at $t=60$ s, resulting in a confined trimer that persists until approximately $t=119$ s. Switching the laser off again at $t=120$ s leads to complete disassembly, with the nanoparticles returning to diffusive motion over the droplet surface by $t=124$ s. The repeated assembly-disassembly cycle demonstrates that the observed nanoparticle structures are dynamically maintained by continuous optical excitation rather than irreversible particle aggregation. In particular, the rapid loss of confinement following laser removal, followed by renewed particle recruitment and assembly upon reillumination, establishes that the interfacial optothermal confinement can be repeatedly switched between assembled and freely diffusing states.

A complementary observation is provided by Supplementary Video 3, which presents a broader field of view of the entire droplet and surrounding nanoparticles. Under laser illumination, multiple nanoparticles are recruited toward the illuminated region and form a localized assembly, whereas switching off the laser results in its disappearance and the recovery of diffusive motion. Although individual nanoparticles cannot be reliably resolved or counted in this wider-field view, the video directly captures the reversible formation and disassembly of the optothermally induced assembly.

The transition between freely diffusing and optothermally confined states is further supported by the mean-squared-displacement (MSD) analysis presented in Supplementary Fig. 4. The freely diffusing nanoparticle exhibits substantially greater spatial exploration, whereas the trapped nanoparticle shows strongly suppressed positional fluctuations. Consistently, the displacement probability distribution is sharply localized for the trapped state compared with the freely diffusing state, providing quantitative evidence for effective optothermal confinement at the illuminated interface. Together, these observations demonstrate that laser illumination provides reversible control over nanoparticle recruitment, confinement, and disassembly, enabling repeated switching between localized collective states and freely diffusing particles at the liquid–liquid interface.

\begin{figure}[h!]
  \centering
  \includegraphics[width=\linewidth]{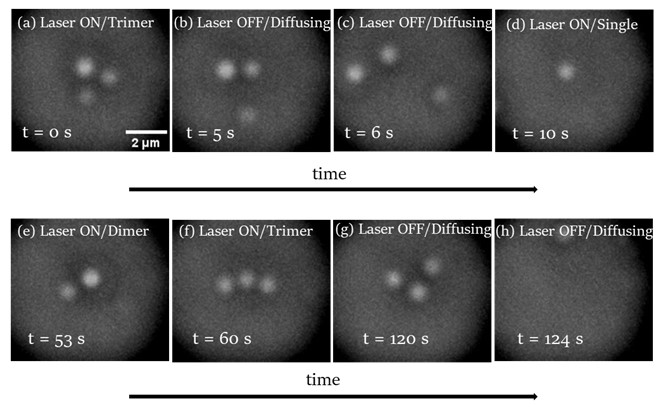}
  \caption{Reversible optothermal assembly and disassembly of plasmonic colloids at a surfactant-stabilized liquid-liquid interface.
  Representative time-series bright-field images showing repeated assembly and disassembly under 532 nm laser illumination. At $t=0$ s, three plasmonic colloids form a confined trimer-like configuration at the illuminated interface. Following laser removal, the colloids progressively diffuse away from the illuminated region, reaching a freely diffusing state by $t=6$ s. Re-illumination at $t=10$ s first confines a single colloid, followed by recruitment of a second colloid at $t=53$ s and formation of a trimer at $t=60$ s. The trimer remains confined until $t=120$ s, after which laser removal leads to progressive disassembly and diffusion over the droplet surface, reaching a dispersed state by $t=124$ s. Supplementary Video 2 shows the complete assembly-disassembly sequence at 2$\times$ playback speeds. Scale bar: 2~$\mu$m.}
  \label{fig:figure03}
\end{figure}

\subsection{ Effect of Plasmonic Colloid Size on Assembly }
To examine the influence of plasmonic colloid size on the characteristic assembly spacing, we performed additional experiments using Au colloids with a diameter of approximately 200 nm. The SEM characterization shown in Supplementary Fig. S5 confirms the approximately spherical morphology and characteristic size of the 200 nm Au colloids used in these experiments. Stable dimer configurations were subsequently observed at the SDS-stabilized BDD-water interface under continuous 532 nm laser illumination. The corresponding two-dimensional particle trajectories and center-to-center separation distribution are presented in Supplementary Fig. S6, demonstrating localized positional fluctuations and a characteristic interparticle spacing for the 200 nm plasmonic colloids. Four independent measurements yielded nearest-neighbor separations of 1.10, 1.17, 1.20, and 1.21 $\mu$m, giving an average separation of $1.17 \pm 0.05~\mu$m. For comparison, seven measurements using 400 nm plasmonic colloids yielded an average separation of $1.21 \pm 0.09~\mu$m. Thus, reducing the colloid diameter from 400 to 200 nm does not produce a systematic shift in the characteristic dimer separation over the investigated size range.
                    
To compare the corresponding optical responses, we calculated the wavelength-dependent absorption and scattering cross sections of the 200 and 400 nm plasmonic colloids using Mie theory (Supplementary Fig. S7). At 532 nm, both sizes exhibit finite absorption and scattering, with substantially larger cross sections for the 400 nm colloids. Therefore, the comparable assembly spacing observed for the two sizes does not simply scale with the magnitude of the single-colloid optical response. Instead, the results suggest that the characteristic $\sim$1.1-1.3 $\mu$m spacing is comparatively insensitive to colloid size under the present interfacial conditions and reflects a characteristic length scale of the interfacial optothermal assembly.  

\subsection{Optothermal Recruitment at the Liquid–Liquid Interface}
To identify the conditions required for recruitment of plasmonic colloids, we performed control experiments in the absence of either the liquid-liquid interface or SDS (Fig. 4). In pure water, the plasmonic colloids exhibited unrestricted Brownian diffusion under continuous 532 nm illumination, with no sustained trapping or directional migration (Fig. 4a; Supplementary Video 4). Similarly, addition of 3 mM SDS in the absence of the liquid-liquid interface did not produce sustained confinement or assembly (Fig. 4b; Supplementary Video 5). Introducing the BDD-water interface without SDS also did not result in stable assembly, even when the laser focus was positioned at the interface (Supplementary Video 6). These controls demonstrate that neither optical illumination, SDS, nor the liquid-liquid interface alone is sufficient to produce sustained optothermal recruitment and confinement.

In contrast, when both the SDS-stabilized liquid-liquid interface and 532 nm illumination were present, a qualitatively different behavior was observed (Fig. 4c). A plasmonic colloid initially undergoing Brownian motion at the interface became confined at the illuminated region. Subsequently, a second colloid located approximately 3~$\mu$m away migrated toward the confined colloid and became incorporated into the assembly at $t=3$ s (Supplementary Video 7). The corresponding trajectories reveal a transition from Brownian diffusion to directed radial migration followed by localized interfacial confinement. Recruitment over several micrometres indicates that the observed particle transport extends beyond the spatial range expected for local optical trapping alone.

The observed trajectories also provide constraints on the underlying transport mechanism. A simple outward Marangoni flow from a heated region is inconsistent with the observed inward radial migration, while isotropic thermophoresis alone does not account for the complete trajectories. In some cases, colloids approaching the illuminated region subsequently leave the focal plane, suggesting an out-of-plane advective component. Similar surface-directed transport followed by out-of-plane motion has been reported for colloids near heated liquid-liquid interfaces and attributed to coupled interfacial optofluidic effects \cite{caciagli2020controlled}. Thermophoretic migration of gold nanostructures under laser heating has also been reported in homogeneous solutions \cite{makihara2019thermophoresis}, while the thermophoretic response depends on particle surface properties and interfacial boundary conditions \cite{burelbach2017colloidal}. In the present system, the SDS-stabilized BDD-water interface further modifies the thermal and hydrodynamic environment. We therefore interpret the recruitment as arising from a coupled interfacial optothermal transport process involving thermal, hydrodynamic, and surfactant-mediated effects, without assigning the motion to a single dominant mechanism. Together, the control experiments and particle trajectories demonstrate that sustained recruitment requires the combined presence of the liquid-liquid interface, SDS, and optical excitation. The resulting interfacial optothermal environment provides the long-range transport and local confinement necessary for the sequential assembly of plasmonic colloids described above.

\begin{figure}[h!]
  \centering
  \includegraphics[width=\linewidth]{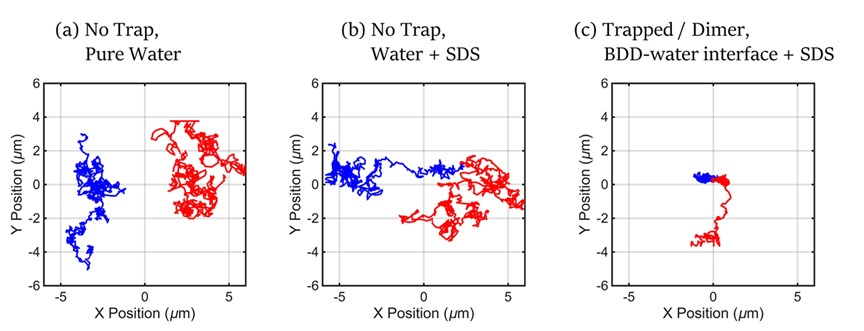}
  \caption{Control experiments demonstrating the conditions required for optothermal recruitment and assembly of plasmonic colloids.
  Representative particle trajectories obtained under different experimental conditions, with the laser focused at the position corresponding to $(x,y)=(0,0)$ in each trajectory plot. (a) In pure water, plasmonic colloids undergo unrestricted Brownian diffusion under continuous 532 nm laser illumination, with no sustained confinement (Supplementary Video 4). (b) In 3 mM SDS solution without a liquid-liquid interface, the colloids similarly exhibit Brownian diffusion without stable assembly under laser illumination (Supplementary Video 5). (c) At the SDS-stabilized BDD-water interface, a colloid is recruited toward the illuminated region and becomes confined, followed by recruitment of a second colloid to form a dimer (Supplementary Video 7). The corresponding trajectories illustrate the transition from diffusive motion to directed recruitment and localized interfacial confinement.}
  \label{fig:figure04}
\end{figure}

\begin{figure}[h!]
\centering
\includegraphics[width=\linewidth]{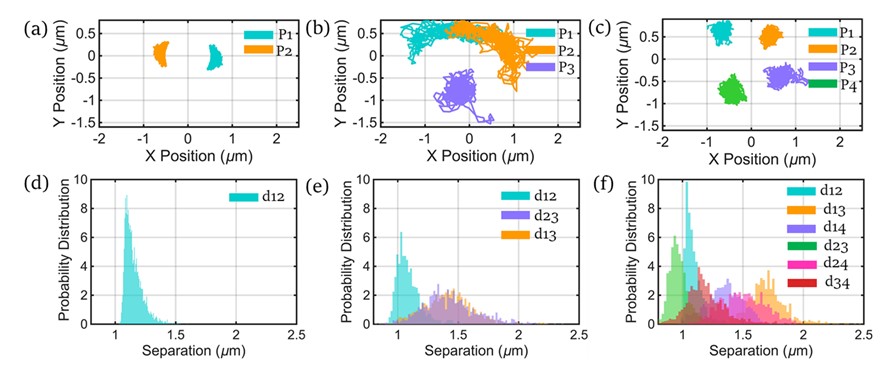}
\caption{Structural dynamics and characteristic interparticle spacing of evolving plasmonic-colloid assemblies.
(a-c) Representative trajectories of plasmonic colloids in dimer (a), trimer (b), and quadramer (c) configurations under continuous 532 nm laser illumination. Each trajectory corresponds to an individual colloid tracked over time, illustrating positional fluctuations and structural reorganization within the assemblies. (d-f) Normalized probability distributions of pairwise center-to-center separations for the corresponding dimer (d), trimer (e), and quadramer (f) configurations. The dimer exhibits a characteristic separation of approximately 1.15~$\mu$m. For the trimer, the nearest-neighbor separation is approximately 1.1~$\mu$m, while the larger separations of 1.41 and 1.43~$\mu$m correspond to non-nearest-neighbor particle pairs. The quadramer exhibits multiple pairwise distance distributions, with nearest-neighbor separations remaining within a comparable micrometre-scale range. The persistence of a characteristic nearest-neighbor spacing during assembly indicates a preferred length scale for the interfacial organization of the plasmonic colloids.}
\label{fig:figure05}
\end{figure}

\subsection{Quantitative Dynamics of Evolving Plasmonic-Colloid Assemblies}
To quantify the evolution of the assembled plasmonic colloids, we analysed particle trajectories and interparticle-distance distributions for dimer, trimer, and quadramer configurations (Fig. 5). The dimer exhibits confined fluctuations around a characteristic interparticle separation of approximately 1.1 to 1.15 $\mu$m, as reflected by the corresponding trajectory and narrow distance distribution. Upon recruitment of a third plasmonic colloid, the dimer reorganises into a trimer configuration while retaining a comparable local separation. In a representative triangular trimer, the distances between the newly recruited colloid and the two original colloids are $d_{13}=1.41~\mu$m and $d_{23}=1.43~\mu$m, while the original dimer separation remains approximately 1.1 $\mu$m. This produces an approximately isosceles triangular configuration, demonstrating that particle addition is accompanied by structural reorganisation rather than simple particle-particle contact.

Further recruitment produces a four-colloid assembly with additional structural reorganisation. A representative configuration exhibits nearest-neighbour separations of $d_{12}=1.10$, $d_{23}=1.00$, $d_{34}=1.19$, and $d_{41}=1.34~\mu$m, while the diagonal distances are $d_{13}=1.69$ and $d_{24}=1.50~\mu$m. The resulting geometry is distorted rhombus-like quadrilateral and noncompact, with the larger diagonal distances distinguishing it from a closely packed aggregate. Thus, successive incorporation of plasmonic colloids leads to the occupation of additional spatial positions while preserving micrometre-scale interparticle separations.

The interparticle-distance distributions further show that the nearest-neighbour separation remains within a comparable range as the assembly evolves from dimer to higher-order configurations. This persistence indicates a characteristic length scale for local organisation under the interfacial optothermal conditions. The trajectories additionally reveal continuous positional fluctuations and structural rearrangement within the assemblies, demonstrating that the observed configurations are dynamically maintained rather than static structures. Together, the trajectory and distance analyses establish an evolutionary assembly pathway in which sequential recruitment is accompanied by progressive many-body reorganisation while maintaining a characteristic interparticle separation. 
 
\subsection{Conclusion}
In conclusion, we demonstrate surfactant-assisted optothermal binding and reversible assembly of plasmonic gold nanoparticles at a liquid–liquid interface under low-power, continuous 532 nm laser illumination. In contrast to conventional scattering-mediated optical binding, this approach exploits the absorption-related optothermal response of plasmonic colloids to drive their recruitment and organization at an illuminated droplet interface. The assemblies evolve sequentially from a single trapped particle to dimers, trimers, quadramers, and occasionally pentamers, while remaining dynamically mobile and maintaining micrometre-scale interparticle separations. 

Laser on/off experiments confirm the reversible nature of the binding process, with disassembly and recovery of Brownian diffusion upon removal of optical excitation. Control experiments establish that sustained particle recruitment requires the combined presence of the SDS-stabilized liquid–liquid interface and optical excitation. The observed trajectories indicate long-range directed transport toward the illuminated region, consistent with coupled interfacial optothermal, thermophoretic, and hydrodynamic effects\cite{caciagli2020controlled}.

Importantly, comparable nearest-neighbour separations for 200 and 400 nm gold nanoparticles indicate that the characteristic spacing is largely insensitive to particle size over the investigated range. Overall, this work establishes surfactant-stabilized liquid–liquid interfaces as a platform for absorption-driven, reversible organization of optothermally bound plasmonic matter and for investigating its spatiotemporal evolution. This approach offers opportunities for studying nonequilibrium colloidal organization at soft interfaces, with potential relevance to nanoparticle assembly on biological surfaces such as cells.

\section*{Supporting Information}
{Use \href{https://drive.google.com/drive/folders/1acC9HzRYZKM7UMHUKDyHae77CrfwFBF-?usp=sharing}{\textbf{this link}} to access the supporting information PDF and video related to the manuscript.}

\section*{Acknowledgements}
Authors thank Ashutosh Shukla, Sneha Boby, Arindam Maity, Simanta Baishya, and Somodip Pal for the fruitful discussion. Sumant Pandey acknowledges the Council of Scientific and Industrial Research-University Grants Commission (CSIR-UGC) for providing a fellowship. This work was funded by the Anusandhan National Research Foundation (ANRF), Grant No. ANRF/ARG/2025/002064/PS, to G. V. Pavan Kumar.

\bibliography{citations}

\end{document}